\documentclass[12pt,preprint]{aastex61}

\shortauthors{Bower et al.}
\shorttitle{Molecular Gas in FRB Host Galaxy}
\begin{document}

\newcommand\degd{\ifmmode^{\circ}\!\!\!.\,\else$^{\circ}\!\!\!.\,$\fi}
\newcommand{\etal}{{\it et al.\ }}
\newcommand{\uv}{(u,v)}
\newcommand{\rdm}{{\rm\ rad\ m^{-2}}}
\newcommand{\msun}{{\rm\ M_{\sun}}}
\newcommand{\msuny}{{\rm\ M_{\sun}\ yr^{-1}}}
\newcommand{\mylesssim}{\stackrel{\scriptstyle <}{\scriptstyle \sim}}
\newcommand{\lsim}{\stackrel{\scriptstyle <}{\scriptstyle \sim}}
\newcommand{\gsim}{\stackrel{\scriptstyle >}{\scriptstyle \sim}}
\newcommand{\sci}{Science}
\newcommand{\sgr}{PSR J1745-2900}
\newcommand{\sgra}{Sgr~A*}
\newcommand{\kms}{\ensuremath{{\rm km\,s}^{-1}}}
\newcommand{\kkms}{\ensuremath{{\rm K\,km\,s}^{-1}}}
\newcommand{\masy}{\ensuremath{{\rm mas\,yr}^{-1}}}
\newcommand{\frb}{FRB 121102}

\def\kbar{{\mathchar'26\mkern-9mu k}}
\def\totd{{\mathrm{d}}}


\title{A Search for Molecular Gas in the Host Galaxy of \frb}
\author[0000-0003-4056-9982]{Geoffrey C.\ Bower}
\affiliation{Academia Sinica Institute of Astronomy and Astrophysics, 645 N. A'ohoku Place, Hilo, HI 96720, USA}
\email{gbower@asiaa.sinica.edu.tw}
\author{Ramprasad Rao}
\affiliation{Academia Sinica Institute of Astronomy and Astrophysics, 645 N. A'ohoku Place, Hilo, HI 96720, USA}
\author{Melanie Krips}
\affiliation{Institut de Radioastronomie Millimetrique (IRAM), Domaine Universitaire, 300 rue de la Piscine, 38406 Saint-Martin-d'Hères, France}
\author{Natasha Maddox}
\author{Cees Bassa}
\affiliation{ASTRON, the Netherlands Institute for Radio Astronomy, Postbus 2, 7990 AA, Dwingeloo, The Netherlands}
\author{Elizabeth A. K. Adams}
\affiliation{ASTRON, the Netherlands Institute for Radio Astronomy, Postbus 2, 7990 AA, Dwingeloo, The Netherlands}
\affiliation{Kapteyn Astronomical Institute, University of Groningen, Postbus 800, 9700 AV, Groningen, The Netherlands}
\author[0000-0002-4119-9963]{C.~J.~Law}
\affiliation{Department of Astronomy and Radio Astronomy Lab, University of California, Berkeley, CA 94720, USA}
\author{Shriharsh P. Tendulkar}
\affiliation{Department of Physics, 3600 University St, Montréal, QC, Canada, H3A 2T8}
\author{Huib Jan van Langevelde}
\affiliation{Joint Institute for VLBI ERIC, Postbus 2, 7990 AA Dwingeloo, The Netherlands}
\affil{Sterrewacht Leiden, Leiden University, Postbus 9513, 2300 RA Leiden, The Netherlands}
\author{Zsolt Paragi}
\affiliation{Joint Institute for VLBI ERIC, Postbus 2, 7990 AA Dwingeloo, The Netherlands}
\author{Bryan~J.~Butler}
\affiliation{National Radio Astronomy Observatory, Socorro, NM 87801, USA}
\author{Shami~Chatterjee}
\affiliation{Cornell Center for Astrophysics and Planetary Science and Department of Astronomy, Cornell University, Ithaca, NY 14853, USA}

\begin{abstract}
We present SMA and NOEMA observations of the host galaxy of \frb\ in the CO 
3-2 and 1-0 transitions, respectively.  We do not detect emission from either transition.  
We set $3\sigma$ upper limits to the CO luminosity 
$L_{CO} < 2.5 \times 10^7\, \kkms {\, \rm pc^{-2}}$ for CO 3-2
and $L_{CO} < 2.3 \times 10^9\, \kkms {\, \rm pc^{-2}}$ for CO 1-0.
For Milky-Way-like star formation properties, we set a $3\sigma$
upper limit on the $H_2$ mass of $2.5 \times 10^8 \msun$, slightly less
than the predictions for the $H_2$ mass based on the star formation rate.
The true constraint on the $H_2$ mass may be significantly higher, however,
because of the reduction in CO luminosity that is common for 
low-metallicity dwarf galaxies like the FRB host galaxy.
These results
demonstrate the challenge of identifying the nature of FRB progenitors through study of the host galaxy molecular gas.  
We also place a limit of 42 $\mu$Jy ($3\sigma$) on the continuum flux density of the
persistent radio source at 97 GHz, 
consistent with a power-law extrapolation of the low frequency spectrum,
which may arise from an AGN or other nonthermal source.
\end{abstract}


\section{Introduction}
Fast radio bursts (FRBs) are millisecond-duration, highly dispersed radio transients \citep{2007Sci...318..777L,2013Sci...341...53T}.  Approximately 30 FRBs have been identified in the past 
decade \citep{2016PASA...33...45P}.  The large dispersion measure (DM) of FRBs suggests 
that they are of extragalactic origin, implying radio luminosities significantly
in excess of any known phenomenon.  A wide variety of models have been
proposed \citep[e.g.,][]{2014A&A...562A.137F,2014MNRAS.439L..46L,2016MNRAS.462..941L,2016MNRAS.458L..19C,2017ApJ...839L...3K}.  These include neutron star progenitors, primarily due to the 
observational similarity between pulsar and FRB properties.  Other 
models include stellar and super massive black holes along with more
exotic high energy phenomena such as cosmic strings \citep[e.g.,][]{2012PhRvD..86d3521C}.

Of the known FRBs, all but one have been observed to produce only a single
pulse in spite of tens to hundreds of hours of follow-up \citep{2015ApJ...807...16L,2015MNRAS.454..457P}.
FRB 121102, on the other hand, has now been detected more than 100 times
\citep{2014ApJ...790..101S,2017ATel.10675...1G}.  This repetition has enabled targeted
observing
campaigns to localize \frb\ to 
an accuracy better than
100 milliarcsec using the Very Large Array 
\citep{2017Natur.541...58C,2017arXiv170507553L}.
and better than 10 mas
using the European Very Long Baseline Interferometry (VLBI) Network
\citep{2017ApJ...834L...8M}.
The FRB is spatially coincident with a compact
($\lsim 1$ mas), persistent radio source that may be associated with an accreting supermassive black hole, a supernova remnant (SNR),
or a pulsar wind nebula (PWN), although each of these identifications is problematic for different reasons.
In particular, the persistent radio source is orders of magnitude more luminous
than any known SNR or PWN, making that explanation difficult.  On the other hand, while the radio
luminosity and upper limit on the X-ray luminosity are consistent with an active
galactic nucleus (AGN), there is
only one known example of a radio luminous AGN in a dwarf galaxy  \citep[Henize 2-10;][]{2011Natur.470...66R}
despite extensive searches \citep[e.g.,][]{2013ApJ...775..116R,2017ApJ...846...44O}.
Further, as noted below, there is no evidence 
of AGN activity in the FRB host galaxy based on optical spectroscopy.

The FRB is associated with a dwarf galaxy at 
$z=0.19273 \pm 0.00008$, 
$<10^8 \msun$
in stellar mass, and diameter $\approx 2\arcsec$ ($\sim 3\ {\rm kpc}$)
\citep{2017ApJ...834L...7T,2017ApJ...843L...8B,2017ApJ...844...95K}.
High resolution imaging with the Hubble Space Telescope (HST) shows a bright, compact (80 mas) knot of
continuum emission coincident with the FRB and the persistent
radio source, along with diffuse, extended emission.

Analysis of emission lines in the 
optical spectrum shows evidence of star formation with
SFR$\approx 0.2 \msuny$ and no evidence for an AGN.
Spectral analysis places a constraint
on the metallicity of $\log_{10} [{\rm O/H}] + 12 \approx 8.0$
\citep{2017ApJ...843L...8B,2004ApJ...617..240K}, indicating significantly sub-solar metallicity,
which is common for dwarf galaxies.  
Low metallicity dwarf
galaxies have been identified as hosts to long gamma-ray bursts (GRBs) and 
superluminous supernova (SLSNe) \citep{2017arXiv170102370M}.

Multi-wavelength studies of GRB hosts have demonstrated the power to constrain
progenitor classes \citep[e.g.,][]{2016SSRv..202..111P}.
ALMA non-detection of carbon monoxide (CO) in two long-duration
GRB hosts has shown an unusually low gas-to-dust ratio at the location
of the GRB, possibly as the result of ultraviolet radiation
dissociation of the CO, providing an important clue to 
the environment from which GRBs originate \citep{2014Natur.510..247H}.
Further, CO spectroscopy can be an important
diagnostic of molecular gas mass and dynamics in 
dwarf galaxies \citep[e.g.,][]{2015Natur.525..218R}.  Finally, in addition to
characterizing star formation in the host galaxy, CO could determine
the dynamical center of the galaxy or detect the presence of an AGN through
high velocity gas \citep{2013Natur.494..328D}.  The CO luminosity for 
low metallicity dwarf galaxies is difficult to estimate and
may be significantly suppressed relative to Milky Way-like systems
\citep{2012AJ....143..138S}.  

In this paper, we report on Submillimeter Array (SMA) and Northern
Extended Millimeter Array (NOEMA)
observations of the host galaxy of \frb\ in CO
transitions of 3-2 and 1-0.  
In Section 2, we describe our observations,
which lead to a non-detection of CO and of the continuum emission at 97 GHz.  
In Section 3, we discuss the significance
of those limits.

\section{Observations and Results}
\subsection{SMA}
Two observing tracks were obtained with the SMA on 17 and 25 December 2016 
for a total of 11 hours on the host galaxy of \frb.  The array was in its compact configuration for both
epochs.  Atmospheric phase fluctuations were large in the first
epoch and the optical depth was high, leaving only a fraction of the data usable.
Weather conditions were better in the second epoch, with 
a typical 225 GHz opacity of 0.05.  In both observations, the receiver was tuned to a frequency
of 289.930 GHz, the rest frequency of the red-shifted CO 3-2 transition.  The SWARM
correlator was configured with a 2.28-GHz bandwidth spectral window containing 
1024 channels (with velocity width $\sim 2 \kms$) centered on the CO 3-2 transition.

Absolute flux calibration was performed using Neptune.  Bandpass calibration was performed using
3C 273 and phase calibration was performed using the compact source J0555+398.  Standard
data reduction techniques using MIR\footnote{\url{https://www.cfa.harvard.edu/~cqi/mircook.html}} and the Multichannel Image Reconstruction, Image Analysis and Dispaly software \citep[MIRIAD;][]{1995adass...4..433S} were performed to flag, calibrate, and image the data.  Good data from the first and second epochs were combined to create the final results.
We obtained a naturally-weighted image with a beam size of $2.5 \times 2.0$ arcsec at position 
angle 69$^\circ$.  

\subsection{NOEMA}
Three observing tracks were obtained with NOEMA on 18 Feb 2017, 9 Apr 2017 and 12 Apr 2017 
for a total of 9.7 hours on source.
NOEMA was in its D configuration with baselines between 16 and 176m.  Receivers were tuned 
to 96.6 GHz for the red-shifted CO 1-0 transition.  
The WideX correlator has a fixed configuration with 3.6GHz wide IF
bands (one per polarisation) and a spectral resolution of 1.95MHz 
 (velocity width $\sim 6\kms$)
covering the CO(1-0) frequency.

Absolute flux calibration was performed on MWC 349 or LkHa 101 and phase calibration was
performed using J0555+398.  Flagging and calibration were performed with 
the Grenoble Image and Line Data Analysis Software \citep[GILDAS;][]{2013ascl.soft05010G} and imaging
was performed in AIPS \citep{2003ASSL..285..109G}.  We obtained a naturally-weighted image with a
beam size of $4.1 \times 3.6$ arcsec at position angle 0$^\circ$.

\subsection{Results}

\begin{figure*}
  \includegraphics[width=0.8\textwidth]{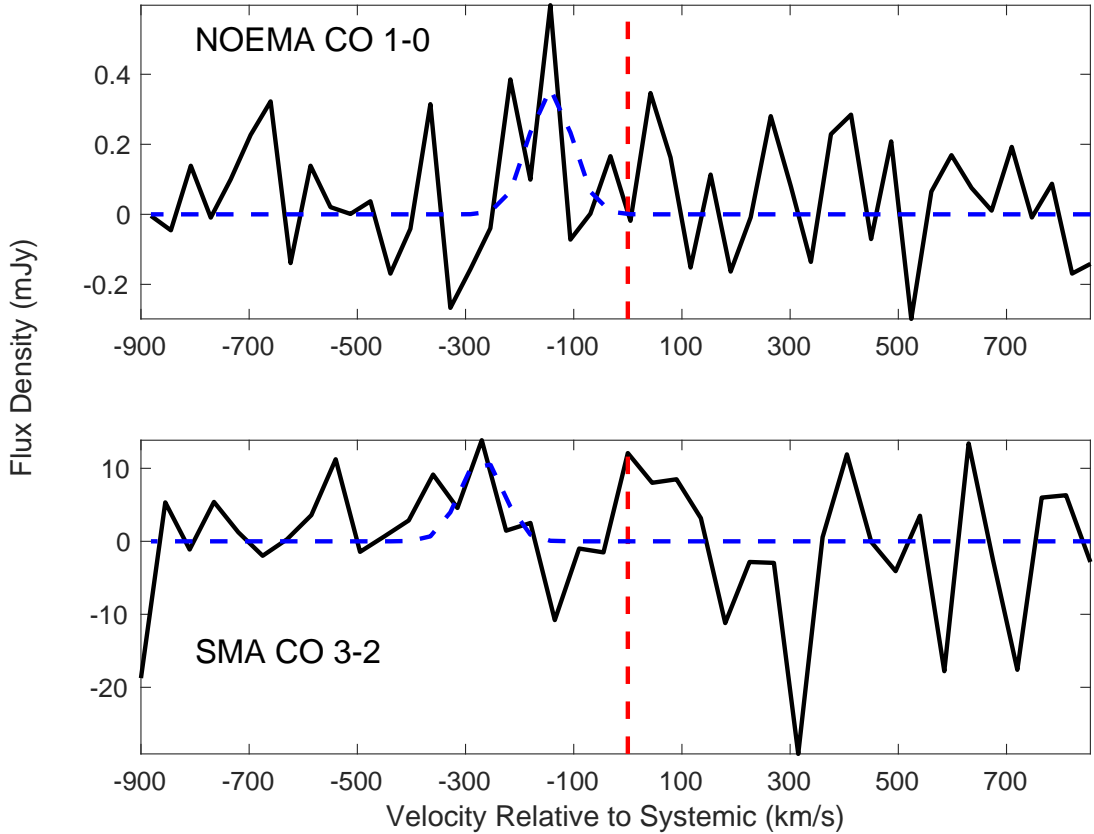}
  \caption{
NOEMA CO 1-0 (top) and SMA CO 3-2 (bottom) spectra in the 1655 $\kms$ centered on the systemic 
velocity.  The spectra are binned into 38 and 45 \kms channels, respectively.  No
emission or absorption lines are present.  
The dashed blue lines show best fit Gaussians with a dispersion
of 40 \kms at the position of the peak intensity.  
\label{fig:noemaspec}}
\end{figure*}

We detect no line source at the position of \frb\ or anywhere else in the SMA and NOEMA maps.  
SMA maps were made with a velocity resolution of 45 $\kms$ achieving an rms of 9 mJy.
NOEMA maps were made with a velocity resolution of 38 $\kms$ achieving an rms of 0.21 mJy.
We also searched the spectra for higher resolution features without any detection.
The spectra are shown in Figure~\ref{fig:noemaspec}.

Molecular gas is typically found 
within a few hundred km/s of the
the systemic velocity for dwarf galaxies.
For example, in a sample of blue compact dwarfs, 
\citet{2016A&A...588A..23A} found CO emission within 200 $\kms$ of
the velocity of the centroid of optical emission lines. 
Velocity offsets and wide lines can differ by hundreds of \kms\
in the case of AGN and ULIRG outflows 
\citep{2014A&A...562A..21C}. 
It is unlikely that such powerful outflows are present in this system:  optical spectroscopy shows 
no evidence for either a supermassive black hole or strong starburst 
activity in the host galaxy. It is possible that the persistent source is the result of a low luminosity
AGN, but such systems do not drive powerful winds.
The redshift uncertainty $\Delta z=8 \times 10^{-5}$ 
corresponds to a velocity uncertainty of 24 $\kms$, less than the width of a single channel.  
We have searched velocities over the range -900 to +800 $\kms$.
Thus, our search over a velocity range of 1700 $\kms$ centered on
the systemic velocity should recover any emission lines present.

We fit Gaussians with a dispersion $\sigma$ of 40 $\kms$ (FWHM $= 2 \sqrt{2  \log{2}} \sigma = 94\, \kms$) at the
position of peak intensity in each spectrum.  As discussed below a dispersion
of $\sim 40 \kms$ is a characteristic width likely to be found in
this system.
We find peak values of
$10.2 \pm 7.2$ mJy and $0.17 \pm 0.14$ mJy
 for the SMA and NOEMA spectra, respectively, both
of which are consistent with non-detections.  
These correspond to $3\sigma$ upper limits
on the integrated line flux densities of
1430 mJy $\kms$
and
21 mJy $\kms$, respectively.
Since Gaussians fitted at the position of the peak intensity of a
noise-like spectrum are a biased
measurement, we also estimate limits biased on the rms noise in the spectra 
integrated over the expected line widths.  
We find $3\sigma$ limits of 1145 mJy $\kms$ and 25 mJy$\, \kms$,
respectively.  
We adopt these latter
values as our upper limits for further discussion.
We also fit Gaussians with an unconstrained width.  These
lead to low significance fits with characteristics dispersions in
the range 400 -- 600 $\kms$ and peak intensities lower by an order of
magnitude than the narrow fits; that is, essentially equivalent integrated
intensity.

We can translate our upper limits on the integrated line flux density
to limits on the CO luminosity\citep{2013ARA&A..51..207B}.  We find 
$3\sigma$ limits of $L_{CO} < 2.3 \times 10^9\, \kkms {\, \rm pc^{-2}}$ for CO 1-0
and $L_{CO} < 2.5 \times 10^7\, \kkms {\, \rm pc^{-2}}$ for CO 3-2.

As discussed in
\citet{2017ApJ...843L...8B}, there is weak evidence for an overdensity of sources near the \frb\ host galaxy 
that could appear through their line emission.  
We searched the full field of view
including an IRAC galaxy 6\arcsec\, southwest of the host galaxy
and did not make any detection of lines at comparable sensitivity to that quoted above.
In the NOEMA band, this corresponds to a redshift window 0.166 to 0.211.

We also constructed a continuum map at 97 GHz from the NOEMA data and detected no sources
in the field above a $3\sigma$ threshold of 42 $\mu$Jy.

\section{Discussion}

In this section, we compare our luminosity limits against theoretical
expectations.  First, we compute an estimate of the expected $H_2$ mass based on
observed star formation rate.  This estimate along with the size of 
the system, lets us set a range of characteristic velocity widths for the
CO line.  Second, we convert our observed CO luminosity limits into 
an $H_2$ mass using Milky-Way-like conversion factors.  Finally, we
consider the impact of non-MW-like conversion factors on the CO luminosity
and on other estimates of the $H_2$ mass.  We also discuss the significance
of the non-detection of continuum emission from the persistent radio source.

The molecular gas mass and dynamics can be estimated using standard relations
derived from Milky Way-like galaxies.  Using 
the average star formation efficiency ${\rm SFE}=5.25 \times 10^{-10} {\rm\, yr^{-1}}$ \citep{Leroy08}, we can
estimate the total molecular gas content from the observed FRB host ${\rm SFR}=0.23 \msuny$ to be
$M_{H_2} = 4 \times 10^8 \msun$.  
This $H_2$ mass is larger than 
the stellar mass estimate of $\sim 10^8 \msun$.
Using a size of $R=1$ kpc, we estimate a velocity
dispersion 
$\sigma \approx \sqrt{GM_{H_2} / R} \approx 40\, \kms$.
We note that dark matter is often dominant in dwarf galaxies so that even in the event of lower
gas or stellar mass, a velocity dispersion in the range of $40 - 50 \kms$ remains reasonable
\citep[e.g.,][]{2015AJ....149..180O}.
This assumed line width is comparable to that seen in dwarf galaxies.
For instance,
\citet{2016A&A...588A..23A} observed dwarf galaxy CO line dispersions in the range of 
18 to 68 $\kms$.

We convert our observed CO luminosity limit into an $H_2$ mass limit
using a MW-like conversion factor for the $H_2-CO$ ratio 
$X_{CO} = 2.3 \times 10^{20} {\rm\, cm^{-2}} (\kkms)^{-1}$.
We use our 40 $\kms$ limits and adopt a ratio $R_{31}=0.5$,  
where $R_{31}$ is
the line ratio between the 3-2 and 1-0 transitions determined by excitation.
$R_{31}$ has an order of magnitude uncertainty.  For a large sample of normal galaxies, 
\citet{2012MNRAS.424.3050W} found $R_{31} = 0.18$.  \citet{2012AJ....143..138S} selected the 2-1/1-0 line ratio $R_{21} = 0.7$ for
their dwarf galaxy sample.  We then estimate $3\sigma$ upper limits to
the $H_2$ mass of 
$10^{10} \msun$ and $2.5 \times 10^8 \msun$ for the 1-0 and 3-2 transitions,
respectively.  The latter value rejects the predicted $4 \times 10^8 \msun$
mass estimate based on the SFR and the assumption of MW-like properties.

The CO luminosity may be significantly suppressed, however, relative to the expectations
for a MW-like system, weakening the constraints that we have on the $H_2$ mass.
Low metallicity has been shown to substantially reduce the CO luminosity, 
possibly as the result of the absence of dust-shielding of UV radiation, 
leading to photodissociation of CO \citep{2013ARA&A..51..207B}.
\frb\ host galaxy has a metallicity $12+\log_{10}{[{\rm O/H}]} = 8.0 \pm 0.1$, 
comparable to the
metallicity of the Small Magellanic Cloud \citep[SMC;][]{1982ApJ...252..461D}.
$X_{CO}$ for the SMC is two orders of magnitude larger than for the MW
\citep{2012AJ....143..138S}.  The Large Magellanic Cloud (LMC) has a metallicity
higher than the SMC by $\sim 0.3$ dex and a value of $X_{CO}$ higher by a factor of 20 times
the MW value \citep{2012AJ....143..138S}.  Results for the 
\citet{2012AJ....143..138S}
sample of
galaxies are consistent with a power-law index $y=-2.0$ to $-2.8$ for
$X_{CO} \propto (12 + \log_{10}{[{\rm O/H}]})^y$, depending on sample
selection criteria.  \citet{2016A&A...588A..23A}
found a value of $y=-1.5 \pm 0.3$ for their sample.  Thus, predictions for 
the \frb\ host galaxy $H_2$ mass have an order of magnitude or more uncertainty.
\footnote{We note that giant molecular clouds (GMCs) in theory
\citep{2010ApJ...716.1191W} and in observation 
\citep{2011ApJ...737...12L} 
indicate lower values for $X_{CO}$ than galaxy-wide measurements
for low-metallicity regions.  
Given that our observations cover the entire galaxy, however, it is
likely that the galaxy-wide conversion factors are more relevant.}

The specific star formation rate (sSFR) and metallicity of the FRB host galaxy are 
typical of the population of low-mass, low-metallicity starforming galaxies known as 
blue compact dwarfs (BCDs) \citep{2016A&A...588A..23A}.
These dwarf galaxies are offset from the Schmidt-Kennicutt relation for star formation, 
and, thus, have shorter gas depletion timescales, or higher SFE. The SFE appropriate 
for the metallicity of the FRB host galaxy is much higher than that for MW-type galaxies. 
If we instead use this higher 
SFE, $\sim 3\times 10^{-8}$, then $M_{H_2} \sim 6.6\times 10^{6}\msun$.
\citet{2016A&A...588A..23A} also find that $X_{CO}$ at low metallicity is a factor of 10 larger than 
for MW-type galaxies. 
Thus, if the \frb\ host galaxy is representative of the BCD population, the combination of reduced $H_2$ mass and increased $X_{CO}$ leads to one to
three orders of magnitude reduction in the CO luminosity relative to the MW estimate.

Dust mass estimates also suggest that the molecular hydrogen density 
could be lower than estimates from the SFR.  
From the H$\alpha$ to H$\beta$ ratio $2.71 \pm 0.26$ \citep{2017ApJ...844...95K}, we estimate an 
upper limit on the dust mass of $\sim 10^5 \msun$ \citep{2013ApJ...763..145D}.  
Stellar population synthesis
fits to the broadband colors of the host galaxy do not require any dust 
contribution, leading to a similarly low estimate for the dust mass
\citep{2017ApJ...843L...8B}.
For a dust-to-gas
ratio of $10^{-2}$, we find an upper limit $M_{H_2} < 10^7 \msun$, more than an order of
magnitude lower than estimated from the 
Milky-Way-like $X_{CO}$, but 
consistent with the low metallicity dwarf estimate.

Our SMA and NOEMA results show, therefore, that the FRB host galaxy is 
underluminous in CO relative to 
galaxies with Milky Way-like star-formation
properties and of comparable star formation rate. 
An order of magnitude or more reduction in
the expected CO luminosity as the result of low metallicity, however, would make the signal
undetectable in either observation.
Deep integrations with ALMA have the potential 
to detect the \frb\ host galaxy if its CO-to-SFR conversion is 
within an order of magnitude of the Milky
Way.

The radio spectrum of the persistent source declines steeply at frequencies
above 10 GHz and includes a 230 GHz non-detection with ALMA at a 50 $\mu$Jy
threshold \citep[$3\sigma$;][]{2017Natur.541...58C}.  
An extrapolation of the low frequency spectrum ($S \propto \nu^{-1}$) 
predicts a 97 GHz flux density of $\sim 20 \mu$Jy. This flux density
is comparable to the $\sim 1\sigma$ threshold in the NOEMA data, and,
therefore undetectable.
Thus, the spectrum of the persistent source
is consistent with that of an optically-thin nonthermal source.
AGN and RSNe often show spectra similar to that of the persistent 
source \citep{2010A&ARv..18....1D}.  The nature of
the persistent source remains ambiguous.  While the spectrum is nonthermal, its 
luminosity significantly exceeds that of any Galactic PWN or RSNe.  An AGN interpretation 
of the spectrum and its luminosity is reasonable but is demographically unlikely
given the rarity of radio-loud AGN in dwarf galaxies.

The localization of \frb\ to a low metallicity dwarf galaxy at cosmological distance 
has provided a substantial clue towards understanding the environment and 
nature of the host object that produces fast radio bursts.  
We have searched for molecular gas emission that would serve as an important diagnostic
of star formation, dynamics, and the presence of an AGN.  
The absence of a CO detection in our observations can be interpreted as
the result of a low $H_2$ mass, a high CO-to-$H_2$ conversion factor, a high
star formation efficiency, or a
combination of these factors.  
If the host galaxy of \frb\ is representative of its class, then 
characterization of other host galaxies through molecular gas 
must rely on discovery of
systems that differ in at least one respect:  higher metallicity,
greater mass, or lower redshift.

\acknowledgements
The Submillimeter Array is a joint project between the Smithsonian Astrophysical Observatory and the Academia Sinica Institute of Astronomy and Astrophysics and is funded by the Smithsonian Institution and the Academia Sinica.
This work is based on observations carried out under project number E16AF001 with the IRAM NOEMA Interferometer. IRAM is supported by INSU/CNRS (France), MPG (Germany) and IGN (Spain).

\facilities{SMA, NOEMA}
\software{MIR, MIRIAD, GILDAS, AIPS, MATLAB}


\end{document}